\documentclass[12pt,oneside]{article}
\usepackage{amsmath,amssymb,graphicx,booktabs}
\usepackage{array}
\usepackage[margin=1in]{geometry}
\usepackage{float}
\usepackage{authblk}
\usepackage{xcolor}

\title{\bf How cell size breaks germ-soma symmetry in a gene regulatory network: a mathematical model of \textit{Volvox carteri} cell differentiation}
\author[1*]{Yuanxiao Gao}
\author[1]{Tian Zhang}
\author[1]{Leijing Qiu}
\author[1]{Hui Cao}
\affil[1]{School of Mathematics and Data Science, Shaanxi University of Science and Technology, 710021 Xi'an, China}
\date{}

\begin{document}
\maketitle

\begin{abstract}
\textit{Volvox carteri} has one of the simplest germ-soma systems: large germ cells and small somatic cells arise from a common pool of cells, but how cell size directs these two cell fates remains unclear.
We built a four-variable ordinary differential equation model in which the HSP70A module $H$ drives a cell-size differentiation signal $S$, and $S$ controls a mutually inhibitory pair of fate determinants, \textit{lag} ($L$) and \textit{regA} ($R$), both of which also possess self-activation.
We used bifurcation analysis, Morris screening and Sobol sensitivity analysis, together with simulations constrained by experimental measurements, to study how $S$ controls the germ-soma decision and to predict the \textit{lag} expression ratio.
The model shows that $S$ acts as a bifurcation parameter: below $S_{\mathrm{th}}\approx0.19$ the network has a single stable state with $L\approx R$, whereas above this threshold it has two stable states, one L-dominant and one R-dominant.
Mutual inhibition is required for these two stable fates, whereas self-activation is not needed for bistability but speeds commitment and changes the predicted \textit{lag} ratio when it is asymmetric.
Sensitivity analysis shows that the H/S dynamics and the L/R expression ratios are controlled by largely separate sets of parameters.
Using the observed H peak-to-plateau ratio of approximately 6 and the approximately 1000-fold difference in \textit{regA} expression between the two cell types as constraints, the model predicts a \textit{lag} ratio near 1000 when the two branches are symmetric.
At an asymmetry ratio of 1.5, the predicted median \textit{lag} ratio is approximately 776 for an $\alpha$ asymmetry, 934 for an $\eta$ asymmetry and near 1000 for a $K$ asymmetry.
These results link cell size to the choice between two stable cell fates and provide a quantitative prediction that can be tested once \textit{lag} has been cloned and sequenced.
\end{abstract}

\section*{Author Summary}

Multicellular organisms must turn genetically identical cells into different cell types.
The green alga \textit{Volvox carteri} provides a simple example: large cells become reproductive germ cells, whereas small cells become non-reproductive somatic cells.
Experiments show that cell size is important for this choice, but it is still unclear how a continuous size difference is converted into two stable fates.
We built a mathematical model of the gene network that controls this decision, focusing on two opposing regulators, \textit{lag} and \textit{regA}.
The model shows that the size signal is required to create two alternative cell states; without it, the network cannot produce a stable germ-soma choice.
Mutual inhibition between \textit{lag} and \textit{regA} is also essential, whereas positive feedback within each branch mainly helps cells settle into a fate.
The model further predicts different \textit{lag} expression ratios depending on which part of the network makes the somatic fate stronger.
Because \textit{lag} remains unsequenced, this prediction offers a way to test competing mechanisms once its expression can be measured.
More generally, the work shows how a simple size cue can be translated into stable cell identities.

\section{Introduction}

Cell differentiation is one of the most fundamental processes in multicellular life \cite{newman2020cell}: it divides a genetically identical population of cells into distinct types that perform different tasks, and it thereby creates the division of labour on which complex organisms depend. 
Understanding how a single regulatory network can reliably produce multiple stable cell fates is therefore a central question in developmental biology\cite{kauffman1969metabolic, gardner2000construction}.
Mathematical models provide a natural framework for this question: by representing a gene regulatory network as a nonlinear dynamical system, they relate its regulatory architecture to the number and stability of cell-fate attractors and to the bifurcations that create or eliminate them\cite{kauffman1969metabolic, huang2007bifurcation}.

\textit{Volvox carteri} is an attractive organism for studying this question because it is one of the simplest multicellular organisms, yet it already shows a clear germ-soma split: a small number of large reproductive cells (germ cells) and a much larger number of small non-reproductive cells (somatic cells) coexist in the same colony \cite{kirk2001germ, kirk2001volvox}. 
The genetic program for this germ-soma split has been characterized through developmental genetics \cite{kirk1997genetic}. 
The two cell types arise from asymmetric cell divisions in which blastomeres of different sizes receive different fate information; the GlsA-dependent machinery that links size asymmetry to cell fate has been characterized in \textit{Volvox} \cite{cheng2003glsa}.
Despite its simplicity, the molecular logic that controls this two-cell-type differentiation is still not fully understood \cite{olson2016cooption}.

Known experiments show that this differentiation is controlled by a small set of genes. 
In particular, \textit{regA} is a master somatic-cell repressor whose expression pattern and enhancer/silencer regulation have been characterized \cite{stark2001two}, and a germ-cell determinant called \textit{lag} has been proposed to act upstream of or in parallel with \textit{regA} \cite{kirk2001germ, kirk2005twelve}. 
Cell-type transcriptomes further show that the two cell fates differ dramatically in gene expression \cite{matt2018cell}. 
Comparative genomics indicates that \textit{Volvox} achieved multicellular complexity mainly by co-opting existing regulatory modules rather than by a large expansion of the transcription-factor repertoire \cite{prochnik2010genomic, umen2012genomics}. 
Cell-size information is believed to bias this decision: cells that stay small follow the somatic program, whereas cells that become large follow the germ program \cite{kirk2001germ, matt2018cell, kirk1993cell}.
However, the precise way in which the cell-size signal controls the gene regulatory circuit, and how it converts a continuous size difference into a discrete fate decision, remains unclear.
Moreover,\textit{lag} itself has not yet been fully located and sequenced, so its quantitative behaviour can only be inferred indirectly.

Beyond \textit{Volvox}, evolutionary models show that terminal, irreversible differentiation is shaped by organismal size and cell death in volvocine algae \cite{gao2025stage, gao2025cellDeath, gao2026volvox}.
To address these gaps, we build a mathematical model of the gene regulatory network underlying germ-soma differentiation under asymmetric cell division in \textit{Volvox carteri}.
The model couples the cell-size differentiation signal $S$ with the mutual inhibition between \textit{lag} and \textit{regA}, and analyses how the size signal controls the appearance of two stable cell fates. 
The model also allows us to explain how cell size drives differentiation and to make a quantitative prediction of the \textit{lag} expression ratio between germ and somatic cells.

\section{Model}

\begin{figure}[!htbp]
\centering
\includegraphics[width=\textwidth]{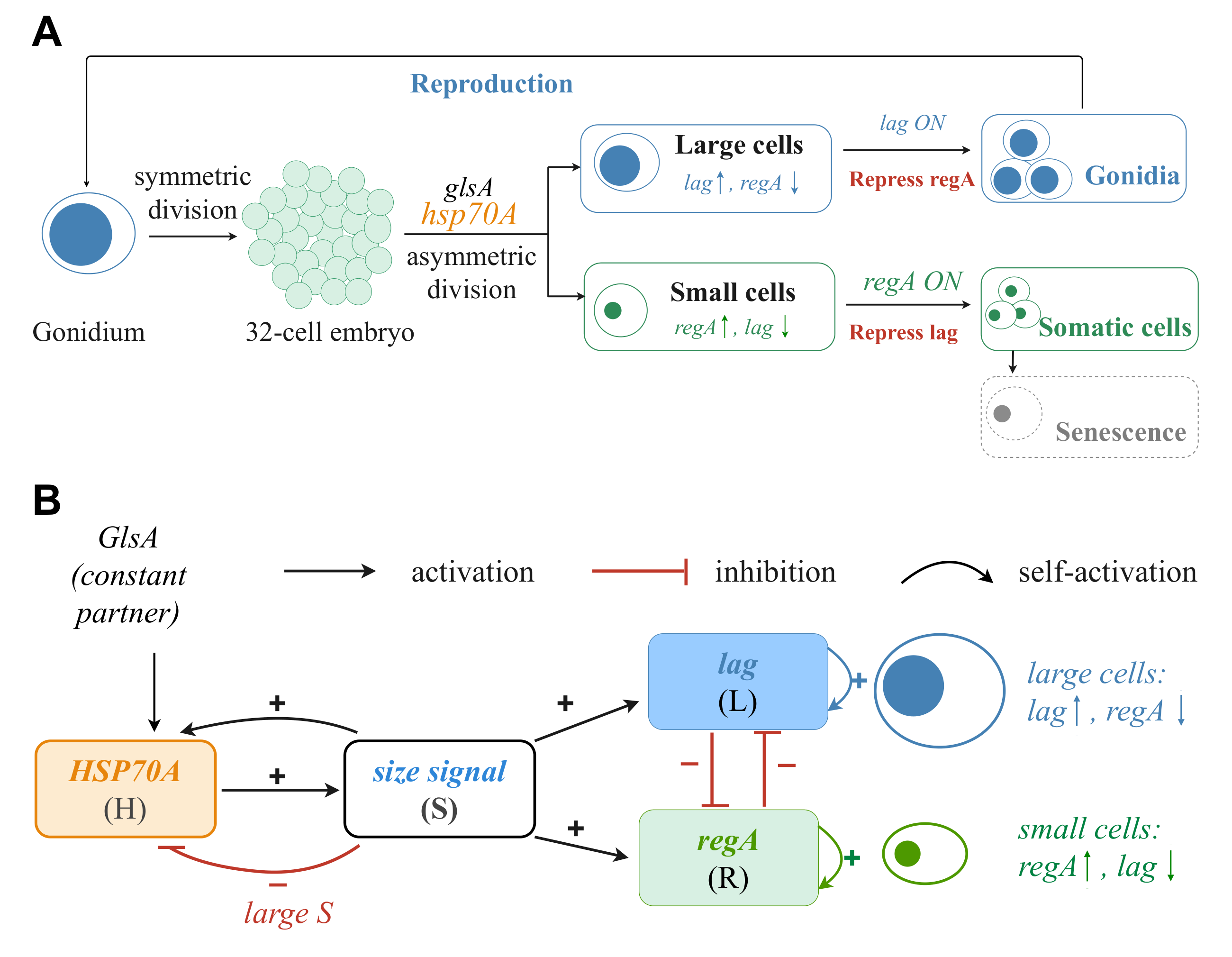}
\caption{Schematic of the \textit{Volvox carteri} life cycle and the gene regulatory network controlling cell differentiation.
(A) Cell differentiation during the \textit{Volvox carteri} life cycle. 
The first five cleavages are symmetric and produce a 32-cell embryo with equal-sized blastomeres.
The sixth cleavage is the first asymmetric division and requires \textit{glsA} and \textit{hsp70A} \cite{cheng2003glsa, cheng2005hsp70a};
this division separates large cells from small cells. 
In small cells, \textit{regA} is switched on, repressing the germ-cell (gonidial) program; these cells become somatic cells and eventually senesce.
In large cells, \textit{lag} is switched on, repressing the somatic program;
these cells become gonidia that retain reproductive potential. 
(B) Schematic of the four-variable ($H$, $S$, $L$, and $R$) gene regulatory network. 
GlsA is the constant partner of the HSP70A module: it provides an activating input to $H$, but is held constant and is not modeled as a dynamic variable.
$H$ and the cell-size signal $S$ activate each other, $S$ activates both fate determinants $L$ and $R$, and $L$ and $R$ mutually inhibit each other while each also self-activates. 
Black arrows indicate activation, red bars indicate inhibition, and curved arrows indicate self-activation; the plus and minus signs mark the sign of regulation and do not encode interaction strength. 
The red bar from large $S$ to $H$ marks the large-$S$ repression term.}
\label{fig:model_schematic}
\end{figure}

\textit{Volvox carteri} develops from a single reproductive cell into a colony containing two cell types: large reproductive germ cells and small terminally differentiated somatic cells \cite{kirk2001germ, matt2018cell} (Fig.~\ref{fig:model_schematic}A).
This binary fate decision is controlled by a small gene regulatory circuit in which the germ determinant \textit{lag} and the somatic repressor \textit{regA} act antagonistically \cite{kirk2001germ, stark2001two, matt2018cell}. 
The expression of these fate determinants is coupled to cell size: cells that remain small activate the somatic program, with high expression of the somatic repressor \textit{regA} ($R$) and low expression of the germ determinant \textit{lag}($L$), whereas cells that enlarge adopt the germ program with high $L$ and low $R$ \cite{kirk2001germ, matt2018cell}. 
Size-dependent asymmetric division is in turn regulated by the J-domain protein GlsA and its partner chaperone HSP70A \cite{cheng2003glsa, cheng2005hsp70a}. 
GlsA provides a relatively constant component of this module, while the periodic activity of HSP70A gates the asymmetric divisions that generate cells of different sizes \cite{cheng2005hsp70a}. 
Based on these observations, we build a minimal model of \textit{Volvox} differentiation consisting of four differential equations for the HSP70A module $H$, the cell-size signal $S$, and the two fate determinants $L$ and $R$.
We adopt a continuous deterministic ordinary differential equation formulation, the standard framework for gene regulatory networks \cite{dejong2002modeling, bolouri2002modeling, karlebach2008modelling}; intrinsic expression noise and parameter fluctuations \cite{lei2009stochastic} are therefore not included here.

The model contains four variables: the master switch HSP70A ($H$), the cell-size differentiation signal $S$, the germ determinant \textit{lag} ($L$), and the somatic repressor \textit{regA} ($R$).
Their regulatory interactions are demonstrated in Fig.~\ref{fig:model_schematic}B and described by the Eq.~(\ref{eq:full_system}).
A small $S$ corresponds to weak size  differentiation (cells still similar in size), and a large $S$ corresponds to strong size differentiation. 
The $H$ module is motivated by the experimental finding that HSP70A and GlsA act as partner chaperones to regulate asymmetric division in \textit{Volvox} \cite{cheng2003glsa, cheng2005hsp70a};
$H$ therefore represents the asymmetric-division module that transmits the size signal to the $L$-$R$ fate switch. 
The $H$ equation therefore contains two opposing $S$-dependent terms. 
The first term activates $H$ when the $S$ is present, representing the onset of asymmetric division. 
The second term represses $H$ at large $S$ through the Hill factor $S^{\,m}/(K_c^{\,m}+S^{\,m})$, consistent with experimental evidence that this module is inhibited by its partner chaperones GlsA and HSP70A \cite{cheng2005hsp70a}. The module is required to initiate asymmetric division but is suppressed once size asymmetry has been established.
We accordingly set $K_c$ larger than $K_{SH}$ so that the activating branch dominates for moderate $S$, while the repressive branch only becomes effective at large $S$.
\begin{equation}\label{eq:full_system}
	\begin{aligned}
		\frac{dH}{dt} &= \alpha_H \cdot \frac{S^{\,n}}{K_{SH}^{\,n} + S^{\,n}}
		\;-\; \gamma_H \cdot H
		\;-\; \delta \cdot \frac{S^{\,m}}{K_c^{\,m} + S^{\,m}} \cdot H \\[10pt]
		\frac{dS}{dt} &= \beta_0 \;+\; \beta_H \cdot \frac{H^{\,n}}{K_{HS}^{\,n} + H^{\,n}}
		\;-\; \gamma_S \cdot S \\[10pt]
		\frac{dL}{dt} &= \beta_L \;+\;
		\frac{S^{\,n}}{K_{SL}^{\,n} + S^{\,n}}
		\left(\alpha_L \;+\; \eta_L \cdot \frac{L^{\,q}}{K_L^{\,q} + L^{\,q}}\right)
		\cdot \frac{K_{RL}^{\,p}}{K_{RL}^{\,p} + R^{\,p}}
		\;-\; \gamma_L \cdot L \\[10pt]
		\frac{dR}{dt} &= \beta_R \;+\;
		\frac{S^{\,n}}{K_{SR}^{\,n} + S^{\,n}}
		\left(\alpha_R \;+\; \eta_R \cdot \frac{R^{\,q}}{K_R^{\,q} + R^{\,q}}\right)
		\cdot \frac{K_{LR}^{\,p}}{K_{LR}^{\,p} + L^{\,p}}
		\;-\; \gamma_R \cdot R.
	\end{aligned}
\end{equation}

The $S$ equation contains three terms: a basal production term $\beta_0$, an $H$-dependent activation term $\beta_H
H^{\,n}/(K_{HS}^{\,n}+H^{\,n})$, 
and a linear decay $-\gamma_S S$. The $H$-dependent term allows asymmetric division to reinforce the cell-size signal, while the decay term keeps the signal bounded.
The $L$ and $R$ equations are structurally almost identical and can therefore be described together. 
Each equation contains four components. The first is a small basal production term ($\beta_L$ or $\beta_R$).
The second is an $S$-dependent production term that combines the maximal $S$-activated production rate
($\alpha_L$ or $\alpha_R$) with the self-activation rate ($\eta_L$ or $\eta_R$).
The third is a mutual-inhibition factor from the opposite gene: $K_{RL}^{\,p}/(K_{RL}^{\,p}+R^{\,p})$ for $L$ and
$K_{LR}^{\,p}/(K_{LR}^{\,p}+L^{\,p})$ for $R$. 
The final term is linear degradation ($-\gamma_L L$ or $-\gamma_R R$).
Because the self-activation terms also depend on $S$, self-maintenance can operate only once the S is strong enough.
The mutual-inhibition factors $K_{RL}^{\,p}/(K_{RL}^{\,p}+R^{\,p})$ and $K_{LR}^{\,p}/(K_{LR}^{\,p}+L^{\,p})$ are written outside the brackets, 
so that cross-repression acts on the entire $S$-dependent production term.
The regulatory functions in Eq.~(\ref{eq:full_system}) are Hill-type terms, which are standard for describing cooperative activation and repression in gene regulatory networks \cite{weiss1997hill, gesztelyi2012hill}. 
The mutual-inhibition core between $L$ and $R$ is the classical genetic toggle switch \cite{gardner2000construction}, a common regulatory motif whose dynamics range from monostable to bistable behavior \cite{tyson2003sniffers}; theoretical studies have long shown that such regulatory circuits can possess multiple stable attractors corresponding to distinct cell types\cite{kauffman1969metabolic}.

\begin{table}[!htbp]
\centering
\caption{Parameter values used in this study.
 The last column gives experimentally supported estimates where available and otherwise the theoretical ranges used for parameter exploration; values retuned for the data-constrained analysis are listed in the table note.}
\label{tab:params}
\footnotesize
\renewcommand{\arraystretch}{1.0}
\begin{tabular*}{\textwidth}{@{\extracolsep{\fill}}>{\raggedright\arraybackslash}m{3.2cm}>{\raggedright\arraybackslash}m{5.4cm}>{\raggedright\arraybackslash}m{1.8cm}>{\raggedright\arraybackslash}m{4.7cm}@{}}
\toprule
Parameter & Description & Value used here & Reference/model range \\
\midrule
$\alpha_H=\alpha_L=\alpha_R$ & Maximal S-activated production rates of $H$, $L$, $R$,
scaled with $\gamma$ ($\alpha/\gamma=10$) & $3.5\,\mathrm{h^{-1}}$
& $\alpha/\gamma$: $0$--$10$;
\cite{gardner2000construction, huang2007bifurcation, huang2024dynamics} \\ \hline
$\eta_L=\eta_R$ & Maximal self-activation rates of $L$, $R$,
scaled with $\gamma$ ($\eta/\gamma=2$) & $0.7\,\mathrm{h^{-1}}$
& $\eta/\gamma$: $0$--$5$;
\cite{huang2007bifurcation, huang2024dynamics} \\ \hline
$K_{SH}=K_L=K_R=K_{SL}=K_{SR}=K_{RL}=K_{LR}$ & Half-saturation constants of
$S$/$H$ and $L$-$R$ regulation & 0.5 & $0.5$--$1.5$;
\cite{huang2007bifurcation, huang2024dynamics} \\ \hline
$K_c$ & Half-saturation, $S$-dependent repression of $H$ & 1.5$^{\dagger}$ & $1.0$--$2.0$;
\cite{huang2007bifurcation, huang2024dynamics} \\ \hline
$\beta_L=\beta_R$ & Basal production rates of $L$, $R$,
scaled with $\gamma$ ($\beta/\gamma=0.01$) & $0.0035\,\mathrm{h^{-1}}$$^{\dagger}$
& small; often $0$ \cite{gardner2000construction}, up to $0.1$
\cite{huang2007bifurcation} \\ \hline
$\gamma_H=\gamma_L=\gamma_R$ & Degradation rates of $H$, $L$, $R$ &
$0.35\,\mathrm{h^{-1}}$ & $0.0007$--$0.35\,\mathrm{h^{-1}}$$^{\ddagger}$;
\cite{mastrobuoni2012proteome, martin2012proteome, raynaud2007mca1, finkelstein2011abi4} \\ \hline
$\delta$ & Strength of $S$-dependent repression of $H$,
scaled with $\gamma$ ($\delta/\gamma=20$) & $7.0\,\mathrm{h^{-1}}$$^{\dagger}$
& $3.5$--$10.5$; no direct Volvox estimate \\ \hline
$p$ & Hill coefficient, $L$/$R$ mutual repression & 3 & $2$--$4$;
\cite{gardner2000construction, weiss1997hill, huang2007bifurcation, huang2024dynamics} \\ \hline
$n$ & Hill coefficient, $S$-dependent regulation & 3 & $2$--$4$;
\cite{gardner2000construction, weiss1997hill, huang2007bifurcation, huang2024dynamics} \\ \hline
$m$ & Hill coefficient, $S$-dependent repression of $H$ & 3$^{\dagger}$ & $2$--$4$;
\cite{gardner2000construction, weiss1997hill, huang2007bifurcation, huang2024dynamics} \\ \hline
$q$ & Hill coefficient, self-activation & 2 & $1$--$3$;
\cite{gardner2000construction, weiss1997hill, huang2007bifurcation, huang2024dynamics} \\ \hline
$\beta_0^{\ast}$ & Basal production rate of the cell-size signal $S$ &
$0.055\,\mathrm{h^{-1}}$$^{\dagger}$ & $0.03$--$0.08$; no direct \textit{Volvox} estimate \\ \hline
$\beta_H^{\ast}$ & Maximal H-dependent reinforcement rate of $S$ &
$0.9\,\mathrm{h^{-1}}$$^{\dagger}$ & $0.5$--$1.3$; no direct \textit{Volvox} estimate \\ \hline
$K_{HS}^{\ast}$ & Half-saturation constant of $S$ activation by $H$ & 0.5$^{\dagger}$ &
$0.3$--$0.8$; \cite{huang2007bifurcation, huang2024dynamics} \\ \hline
$\gamma_S^{\ast}$ & Damping rate of the cell-size signal $S$ &
$0.1\,\mathrm{h^{-1}}$$^{\dagger}$ & $0.05$--$0.20$; no direct \textit{Volvox} estimate \\ \hline
\bottomrule
\end{tabular*}

\vspace{2pt}
\footnotesize\raggedright $^{\ast}$ Parameters marked with an asterisk are used only in the full dynamics and are not used in the clamped-$S$ analysis.
$^{\dagger}$ In the data-constrained analysis (Sec.~\ref{sec:result3}), the values are retuned to $K_c=2.275$, $\delta=7.60$, $m=3.87$,
$\beta_0^{\ast}=0.0523$, $\beta_H^{\ast}=0.634$, $K_{HS}^{\ast}=0.264$,
$\gamma_S^{\ast}=0.116$, and $\beta_L=\beta_R=0.00394$;
all other parameters remain at the values above.
$^{\ddagger}$ The $\gamma$ range is based directly on experimental protein-turnover measurements; the remaining ranges in this column are theoretical ranges from models of bistable gene-regulatory switches, not direct \textit{Volvox} measurements. 
Parameters of the $S$ equation have no direct experimental estimate.
$K_c$ and $K_{HS}$ have no counterpart in the two-gene switch models cited above, whose single shared half-saturation constant is dimensionless; their ranges are set in this study to span the bistable regime, subject to the requirement $K_c>K_{SH}$.
\end{table}

The full model is nonlinear and strongly coupled, so analytical solutions are not practical;
we therefore analyze it mainly through numerical simulation.
Because direct kinetic estimates for \textit{Volvox} \textit{lag}, \textit{regA} and HSP70A are unavailable, the parameter values in Table~\ref{tab:params} combine experimentally constrained estimates with theoretical ranges selected to produce bistable behavior.
Whole-proteome turnover measurements from the close relatives \textit{Chlamydomonas reinhardtii} (median half-life $\approx45$ h, 
$\gamma\approx0.015\,\mathrm{h^{-1}}$) \cite{mastrobuoni2012proteome} and
\textit{Ostreococcus tauri} ($\gamma\approx0.0007$--$0.0058\,\mathrm{h^{-1}}$)
\cite{martin2012proteome} give a bulk algal degradation range of about $0.0007$--$0.015\,\mathrm{h^{-1}}$. 
These bulk values, however, average over all proteins and underestimate the faster turnover of short-lived regulatory
proteins: the \textit{Chlamydomonas} regulator MCA1, for example, has a half-life of about 2 h ($\gamma\approx0.35\,\mathrm{h^{-1}}$) \cite{raynaud2007mca1}, and transcription factors and fate regulators in algae and plants are typically degraded on timescales of minutes to a few hours \cite{finkelstein2011abi4, shen2005pif1, zenser2003auxiaa, li2017degradation}.
Because \textit{lag}, \textit{regA} and HSP70A are such fate regulators and must respond within the developmental schedule of \textit{Volvox} (asymmetric division begins at about 3 h) \cite{kirk2001germ}, we adopt
$\gamma_H=\gamma_L=\gamma_R=0.35\,\mathrm{h^{-1}}$,
the fast end of the regulatory-protein range, rather than the slower bulk-proteome estimate. The production terms are scaled together with
$\gamma$ ($\alpha_H=\alpha_L=\alpha_R=3.5\,\mathrm{h^{-1}}$,
$\eta_L=\eta_R=0.7\,\mathrm{h^{-1}}$,
$\beta_L=\beta_R=0.0035\,\mathrm{h^{-1}}$) 
so that the L-R dimensionless ratios $\alpha/\gamma$, $\eta/\gamma$ and $\beta/\gamma$ are unchanged from
the slow-$\gamma$ set; 
consequently, the L-R steady states do not depend on the absolute value of $\gamma$.
For the H module we set $\delta=7.0\,\mathrm{h^{-1}}$ ($\delta/\gamma=20$), so that the clamped $H^*(S)$ profile is nonmonotonic, with a peak at intermediate $S$ and a lower plateau at large $S$. 
This qualitative shape is motivated by the experimentally observed rise and subsequent decline of the HSP70A-dependent asymmetric-division module \cite{cheng2005hsp70a}.
Because $H(S)$ has not been measured directly, this is used as a qualitative constraint rather than a quantitative fit.
Only the relaxation timescale of the dynamic trajectories is affected by the absolute value of $\gamma$. The small basal rates represent low-level basal expression, the Hill coefficients are kept within measured ranges observed in gene-regulatory switches \cite{weiss1997hill},
and the remaining dimensionless production and saturation parameters follow theoretical ranges from models of bistable gene-regulatory switches \cite{huang2007bifurcation, huang2024dynamics}. 
The model therefore uses experimentally supported values where they are available and otherwise explores clearly stated theoretical ranges.
The parameter values used in the model are summarized in Table~\ref{tab:params}.

Based on this model and parameter selection, we investigate \textit{Volvox} cell differentiation from three perspectives:
first, how the cell-size signal $S$ acts as the bifurcation parameter controlling the bistable L-R fate decision; second, how parameter uncertainty influences differentiation through Sobol sensitivity analysis; 
and third, how experimentally constrained ratios of $H$ and $R$ can be used to predict the likely expression ratio of \textit{lag}.

\section{Results}

\subsection{$S$ is necessary for cell-fate bistability}
\label{sec:result1}

\begin{figure}[!htbp]
\centering
\includegraphics[width=0.95\textwidth]{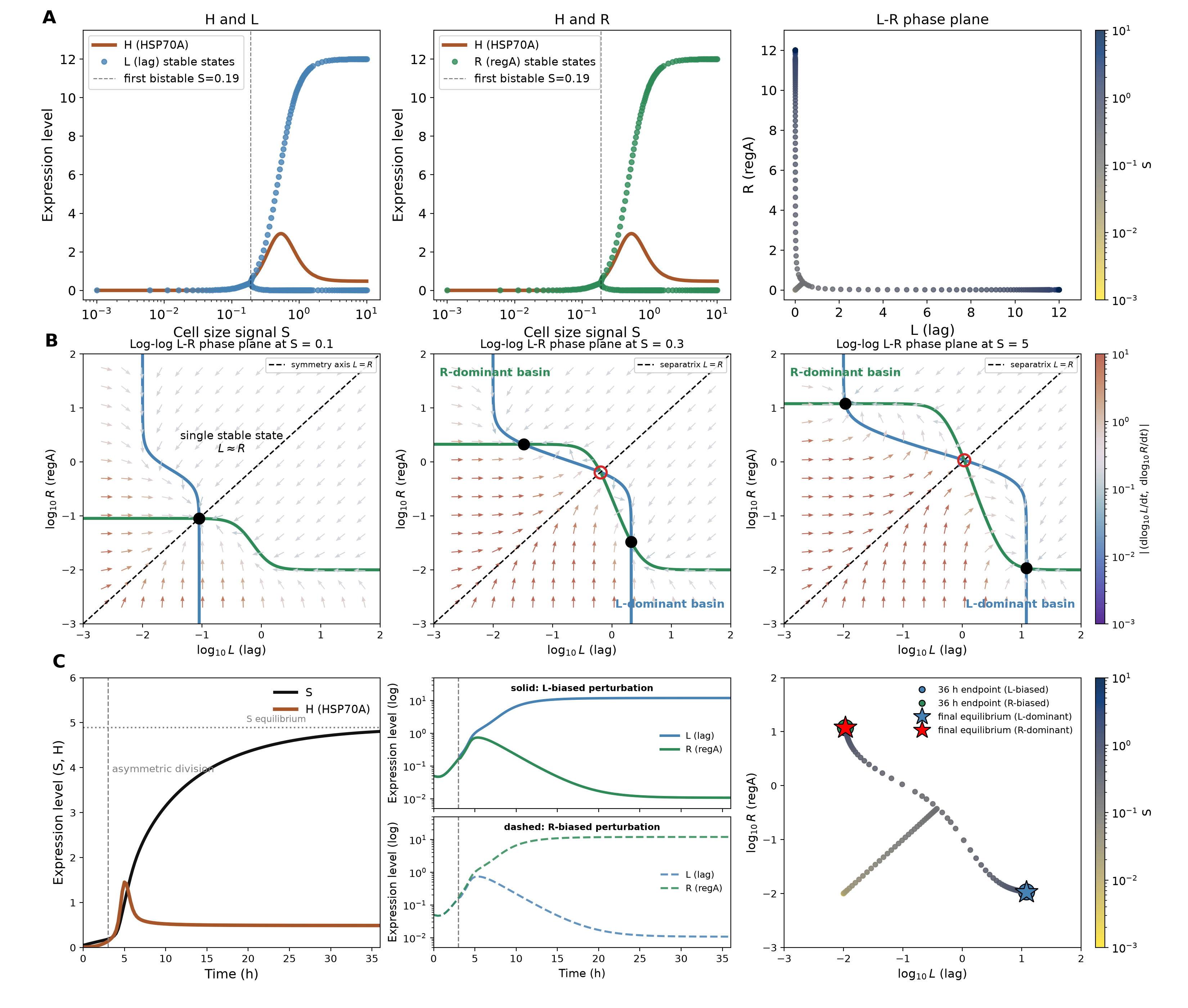}
\caption{Steady-state and phase-plane analysis under clamped $S$, followed by full-dynamics simulations of the self-activation model.
 In panel A, the top row shows the steady-state expression of $H$ together with $L$ and $R$ as functions of the cell-size differentiation signal $S$, and the L-R phase plane in which each stable state is coloured by the corresponding $S$ value. In panel B, the middle row shows the log-log L-R phase portraits at $S=0.1$, $S=0.3$ and $S=5$; black filled circles denote stable steady states and red hollow circles denote saddle points. 
 In panel C, the bottom row shows the full dynamics: the time courses of $S$ and $H$ (left, with the dotted horizontal line marking the full-system equilibrium value of $S$), the time courses of $L$ and $R$ following a small perturbation applied at $t=3$ h from the symmetric value $L=R\approx0.161$, with $L$ raised and $R$ lowered by $0.02$ to $(L,R)\approx(0.181,0.141)$ (L-biased, solid) or the reverse to $(0.141,0.181)$ (R-biased, dashed) (middle), and the 36 h dynamic endpoints overlaid on the clamped-$S$ L-R stable states as a function of $S$ (right); the star symbols mark the final full-system equilibria that the two trajectories approach.
 The full dynamics starts from $(H,S,L,R)=(0.01,0.05,0.05,0.05)$.
All parameters are as in Table~\ref{tab:params} ($\gamma=0.35$~h$^{-1}$, $\alpha=3.5$~h$^{-1}$, $\eta=0.7$,
$\beta=0.0035$, all seven $K=0.5$, $K_c=1.5$, $\delta=7.0$, $p=n=m=3$, $q=2$, and $\beta_0=0.055$, $\beta_H=0.9$, $K_{HS}=0.5$, $\gamma_S=0.1$ for the starred S-equation terms); bistability first appears at $S_{\rm th}\approx0.19$.}
\label{fig:results1}
\end{figure}

In the clamped-$S$ analysis, $S$ is held fixed and all other parameters are taken from Table~\ref{tab:params},
so the $L$-$R$ subsystem is symmetric.
$H^*(S)$ rises to a peak at intermediate $S$ and then declines to a lower plateau at large $S$, reflecting activation of the asymmetric-division module at moderate $S$ and its inhibition once the size asymmetry has been established.
For the L-R subsystem, a small $S$ activates $L$ and $R$ only weakly, so they are produced mainly at their basal rates and their mutual inhibition is too weak to break the $L=R$ symmetry. The system therefore has a single stable state with $L\approx R$.
At $S=0.1$, $L=R\approx0.089$, so the cells remain undifferentiated (Fig.~\ref{fig:results1}A,B). 
As $S$ increases, activation by $S$ approaches saturation and raises production of both genes, strengthening their mutual inhibition. The symmetric $L\approx R$ state then loses stability, and bistability first appears at $S_{\mathrm{th}}\approx0.19$.
Above the threshold, two stable steady states coexist, one L-dominant ($L$ large, $R$ small) and one R-dominant ($R$ large, $L$ small). 
At $S=0.3$, the two stable states are
$(L,R)\approx(2.12,0.033)$ and $(0.033,2.12)$, separated by a saddle point on the diagonal $L=R$ at $(0.65,0.65)$; at $S=5$, they are $(L,R)\approx(12.0,0.011)$ and $(0.011,12.0)$, with the saddle at $(1.07,1.07)$ (Fig.~\ref{fig:results1}A,B). 
Once bistable, a larger $S$ raises the production of whichever gene is already high while the stronger mutual inhibition further suppresses its partner, so the two stable states move away from the diagonal and the L-R expression difference grows with the cell-size signal.
The diagonal $L=R$ is the separatrix between the two basins of attraction: any state with $L>R$ converges to the L-dominant state, and any state with $L<R$ converges to the R-dominant state.

In the full dynamics, $S$ is not constant: starting from a small value, it crosses the bistability threshold around $t\approx3$ h, the time of the first asymmetric divisions in \textit{Volvox} \cite{kirk2001germ}, and continues to increase during development (Fig.~\ref{fig:results1}C). 
The parameters of the $S$ equation are chosen so that $S$ crosses $S_{\rm th}$ at this stage, thereby anchoring the simulated onset of the L-R divergence to the observed timing of asymmetric division.
To represent the fate bias imposed by this first asymmetric division, we apply a small symmetric perturbation to $L$ and $R$ at $t=3$ h and follow the two resulting trajectories. 
The $L$ and $R$ trajectories remain nearly symmetric before the threshold and begin to diverge after it, eventually reaching the L-dominant or R-dominant stable states predicted by the clamped analysis within about 36 h. 
The full dynamics therefore confirms that the clamped-$S$ branches are the attracting states of the system and that differentiation begins only after $S$ has crossed $S_{\rm th}$.

Taken together, these analyses show that the cell-size signal $S$ is a necessary condition for cell-fate bistability: below $S_{\rm th}\approx0.19$ the system cannot generate distinct cell fates, and in the full dynamics $L$ and $R$ begin to separate only after $S$ crosses the threshold.
A large $S$ alone, however, is not sufficient, because mutual inhibition between $L$ and $R$ is also required; removing the cross-inhibitory terms leaves $L$ and $R$ independent with a single $S$-dependent steady state each, so bistability cannot occur at any $S$. 
Self-activation, in contrast, is not required for bistability and only accelerates differentiation: without the self-activation terms the bistable window persists, but the two committed fates are approached more slowly (Figure S1, Supplementary Information).

\subsection{Sensitivity analysis of model parameters}
\label{sec:result2}

Our goal is to test whether the model can reproduce the observed growth and differentiation of \textit{Volvox carteri}. 
Among the model parameters, only the degradation rates are fixed at $\gamma=0.35\,\mathrm{h^{-1}}$ from the protein-turnover data; the remaining 14 parameter combinations (Table~\ref{tab:params}) are not directly constrained by experiments.
In the sensitivity analysis, we separate $\alpha_H$ from $\alpha_L$ because they enter different equations:
$\alpha_H$ controls $H$ production, whereas $\alpha_L$ and $\alpha_R$, which we set equal here, control production of $L$ and $R$.
Setting $\alpha_R=\alpha_L$ keeps the L-R subsystem symmetric, consistent with the analysis in Sec.~\ref{sec:result1}, where $S_{\mathrm{th}}$ is defined by the loss of stability of the symmetric state $L\approx R$. 
This gives 15 screened parameters. The effect of an $\alpha_R\neq\alpha_L$ asymmetry is examined separately in Sec.~\ref{sec:result3}.
Systems-biology models often show uneven parameter sensitivities: a few parameter combinations control most of the output behavior, while many others have little effect \cite{gutenkunst2007universally}.
It is therefore useful to determine which parameters have strong effects and which can be held at their reference
values.
Because the model has many parameters, we first apply the Morris screening method \cite{morris1991factorial} and then
use Sobol indices to estimate the contributions of the retained parameters \cite{sobol2001global}.

We examine four outputs.
The first is the bistability threshold $S_{\mathrm{th}}$, defined by the loss of stability of the symmetric state $L\approx R$; its sensitivity shows which parameters control where the fate switch occurs.
The second is the peak-to-plateau ratio $H_{\mathrm{peak}}/H_{\mathrm{plateau}}$ of the full $H(t)$ trajectory.
Here $H_{\mathrm{plateau}}$ is defined as the late-stage H level at 36~h, a time point at which the L-R fate decision has approached its stable state in the model.
We use this ratio to test whether the model captures the observed rise and later decline of the HSP70A-dependent asymmetric-division module \cite{cheng2005hsp70a}.
The third is the L-R expression difference $\log_{10}(L_{\mathrm{high}}/R_{\mathrm{low}})$, which indicates how far $L$ and $R$ have separated in the committed state, and the fourth is the \textit{regA} ratio $\log_{10}(R_{\mathrm{high}}/L_{\mathrm{low}})$; in the data, \textit{regA} expression differs by about 1000-fold between the two cell types \cite{matt2018cell}, which provides an experimental constraint on this ratio.
Because the $L$ and $R$ equations are symmetric, $\log_{10}(R_{\mathrm{high}}/L_{\mathrm{low}})=\log_{10}(L_{\mathrm{high}}/R_{\mathrm{low}})$ and $\log_{10}(R_{\mathrm{high}}/R_{\mathrm{low}})=\log_{10}(L_{\mathrm{high}}/L_{\mathrm{low}})$.
Morris screening and Sobol sensitivity analysis used the same output definitions: $S_{\mathrm{th}}$ and the L-R expression ratios were computed from clamped-$S$ steady states, whereas the H peak-to-plateau ratio was computed from the full four-variable dynamics.
In the Sobol analysis, the two L-R expression ratios were evaluated at a clamped value of $S$ chosen separately for each parameter sample.
We used $S_{\mathrm{ref}}=\min(3S_{\mathrm{th}},0.6S_{\mathrm{hi}})$, where $S_{\mathrm{th}}$ and $S_{\mathrm{hi}}$ are the lower and upper ends of the bistable interval.
This places the evaluation point at a representative intermediate value inside each sample's bistable range, rather than at a single absolute value of $S$.
Because parameter sensitivities can vary with the chosen value of $S$, we also examined this dependence within the bistable range; the results are reported in the Supplementary Information.

\begin{figure}[!htbp]
\centering
\includegraphics[width=\textwidth]{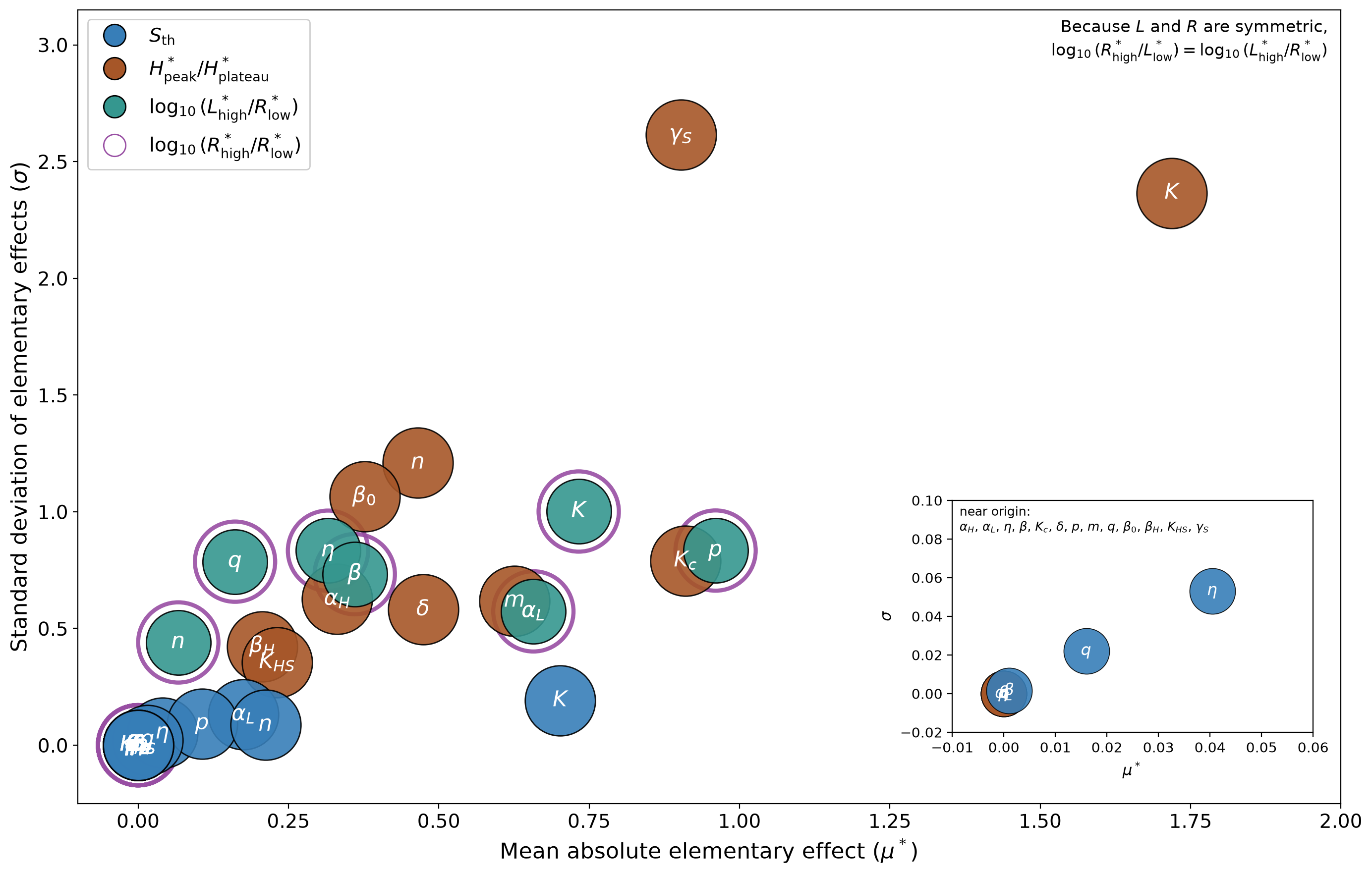}
\caption{Morris screening results.
	Each circle is one parameter and is labelled with its symbol.
	The legend identifies the four outputs.
	The inset enlarges the region $\mu^*\le0.06$ and $\sigma\le0.1$, and lists parameters with $\mu^*\le0.005$ and $\sigma\le0.005$ as near the origin.
	The screening used $k=15$ input parameters, $r=120$ trajectories, and $\ell=4$ levels.}
\label{fig:morris}
\end{figure}

The Morris method estimates, for each parameter, the elementary effect along random trajectories:
\begin{equation}
EE_i^{(j)} = \frac{f(x_1,\ldots,x_i+\Delta,\ldots,x_k) - f(x_1,\ldots,x_k)}
{\Delta},
\end{equation}
where $f$ is the model output, $k$ is the number of input parameters, $x_i$ is the $i$-th parameter normalized to $[0,1]$, and $j$ indexes one of the $r$ random trajectories; each trajectory perturbs every parameter once.
The step $\Delta$ is fixed, typically $\Delta=\ell/[2(\ell-1)]$ with $\ell$ levels, chosen so that $x_i\pm\Delta$ remains within $[0,1]$ and the perturbed point stays on the grid. 
We summarize these effects by
\begin{equation}
\mu_i^* = \frac{1}{r}\sum_{j=1}^{r}\left|EE_i^{(j)}\right|, \qquad
\sigma_i = \sqrt{\frac{1}{r-1}\sum_{j=1}^{r}
\left(EE_i^{(j)}-\mu_i\right)^2},
\end{equation}
with $\mu_i = \frac{1}{r}\sum_{j=1}^{r}EE_i^{(j)}$, where $r$ is the number of trajectories.
Large $\mu^*$ indicates a strong overall influence, while large $\sigma$ indicates nonlinearity or interactions. 
Morris screening shows that $S_{\mathrm{th}}$ depends on the L-R parameters $\alpha_L$, $\eta$, $\beta$, $K$, $p$, $n$ and $q$.
The full-dynamics H peak-to-plateau ratio depends mainly on the H/S parameters $\alpha_H$, $K$, $K_c$, $\delta$, $n$, $m$, $\beta_0$, $\beta_H$, $K_{HS}$ and $\gamma_S$.
The same L-R parameters also control the two expression outputs $\log_{10}(L_{\mathrm{high}}/R_{\mathrm{low}})$ and $\log_{10}(R_{\mathrm{high}}/R_{\mathrm{low}})$ (Fig.~\ref{fig:morris}).
Here $K$ denotes the shared half-saturation constant of the Hill terms (Table~\ref{tab:params}), and is distinct from $K_c$, the half-saturation constant of the $S$-dependent repression of $H$.
Parameters with small $\mu^*$ and small $\sigma$ are fixed at their reference values (Fig.~\ref{fig:morris}).
Morris screening and the Sobol analysis used the same parameter ranges.
For $\alpha_H$, $\alpha_L$, $\eta$, $\beta$ and $\delta$, each parameter was varied by $\pm50\%$ around its reference value.
All other parameters were varied over the ranges listed in Table~\ref{tab:params}.
We use Morris to identify, over all fifteen screened parameters, which ones can change the outputs, and Sobol to estimate how much the retained parameters contribute near the reference values.

Based on the Morris screening, we retain the influential parameters and compute the first-order and total-order Sobol indices
\cite{sobol2001global}:
\begin{equation}
S1_i = \frac{\mathrm{Var}_{x_i}\left(E_{x_{-i}}(f\,|\,x_i)\right)}
{\mathrm{Var}(f)}, \qquad
ST_i = 1 - \frac{\mathrm{Var}_{x_{-i}}\left(E_{x_i}(f\,|\,x_{-i})\right)}
{\mathrm{Var}(f)}.
\end{equation}
Here $f$ is the model output, $x_i$ is the value of parameter $i$, and $x_{-i}$ denotes all parameters except $x_i$.
$E_{x_{-i}}(f\,|\,x_i)$ is the mean of $f$ over all parameters except $x_i$ while keeping $x_i$ fixed.
The resulting mean depends only on $x_i$, and $\mathrm{Var}_{x_i}(E_{x_{-i}}(f\,|\,x_i))$ is the variance of this mean as $x_i$ varies.
$S1_i$ is this variance divided by $\mathrm{Var}(f)$, the total output variance when all parameters vary; it therefore measures the part of the output variance explained by the main effect of $x_i$ alone.
The conditional mean $E_{x_{-i}}(f\mid x_i)$ is the best mean-square predictor of $f$ that depends on $x_i$ alone.
Its variance is therefore the part of $\mathrm{Var}(f)$ that can be attributed to the main effect of $x_i$.
Likewise, $E_{x_i}(f\,|\,x_{-i})$ is the mean of $f$ over $x_i$ while all other parameters are held fixed.
$ST_i$ is one minus the variance of this conditional mean, normalized by $\mathrm{Var}(f)$; it measures the main effect of $x_i$ together with all interactions involving $x_i$.
For each parameter, a small difference between $S1_i$ and $ST_i$ indicates that its main effect dominates, whereas a large difference indicates that interactions involving that parameter contribute substantially to the output variance.
The total variance $\mathrm{Var}(f)$ and the conditional variances are estimated from the Monte Carlo sample using Saltelli's resampling design \cite{saltelli2010variance}; further implementation details are given in the Supplementary Information.

Our results show that $K$ is the only parameter that affects all four outputs, and it accounts for most of the variation in $S_{\mathrm{th}}$ and in the H peak-to-plateau ratio.
Here, $K$ is the shared value of the seven half-saturation constants in Table~\ref{tab:params}, so its index reflects simultaneous changes in all seven thresholds.
For $S_{\mathrm{th}}$, $S1\approx0.8$ and $ST\approx0.83$, with smaller contributions from $n$ and $\alpha_L$.
The small gap between $S1$ and $ST$ indicates that $K$ affects $S_{\mathrm{th}}$ mainly through its own main effect.
For the full-dynamics H peak-to-plateau ratio, $K$ is again the strongest parameter ($S1\approx0.5$ and $ST\approx0.64$), followed by $K_c$, $m$, $\gamma_S$ and $n$.
The larger gap between $S1$ and $ST$ here points to a stronger interaction contribution, which is consistent with $K$ entering several nonlinear Hill terms in the H/S subsystem.
For $S_{\mathrm{th}}$, the retained first-order indices sum to $0.98$ and the corresponding total-order indices sum to $1.02$, so the two sums are close and the response is mainly additive.
For the H peak-to-plateau ratio, the first-order sum is only $0.75$, whereas the total-order sum reaches $1.33$; this larger gap indicates a much stronger contribution from interactions among parameters.
The L-R expression ratio is controlled most strongly by $p$ ($S1\approx0.47$ and $ST\approx0.52$), followed by $\alpha_L$ ($ST\approx0.18$), $n$ ($ST\approx0.16$) and $K$ ($ST\approx0.14$).
The near equality of $S1$ and $ST$ for $p$ shows that its effect is mostly direct, whereas the self-activation parameters $\eta$ and $q$ account for less than $2\%$ of the variance in the L-R expression ratio.

To identify which individual threshold produces the response to $K$, the insets vary the seven half-saturation constants separately while holding all other parameters at their reference values.
Because only these seven thresholds are varied, the inset indices are conditional on the remaining parameters and are normalized separately from those in the main panels.
They therefore show which threshold controls each output, but they do not provide a numerical decomposition of the main-panel $K$ index.
Under this setting, $K_{SR}$ and $K_{SL}$ dominate $S_{\mathrm{th}}$ ($S1\approx0.22$ and $ST\approx0.63$ and $0.62$, respectively), $K_{SH}$ alone controls the H peak-to-plateau ratio ($S1=ST\approx1.00$), $K_{LR}$ dominates and $K_{RL}$ dominates ($S1\approx0.79$ and $ST\approx0.83$, respectively).
The self-activation thresholds $K_L$ and $K_R$ have negligible effects on all four outputs. 
When $K$ is treated as a single shared parameter, its total-order index for $S_{\mathrm{th}}$ is $0.83$, and the small difference from its first-order index indicates that its effect is mainly direct.
When the seven thresholds are varied independently, their first-order indices sum to approximately $0.51$ and their total-order indices to $1.62$.
The large increase in the total-order sum indicates that interactions become much more important once the thresholds vary independently.
For $K_{SL}$ and $K_{SR}$, in particular, the interaction contribution is larger than their individual main effect.
These patterns are consistent with the model structure: $K$ and $n$ set where the Hill-type terms switch on and how steeply they change; $p$ controls the steepness of mutual inhibition, and $\alpha_L$ sets the maximal $S$-activated production, so together they determine how far the two branches separate.
The parameters $\eta$ and $q$ mainly affect how quickly a branch is reached rather than how far apart the branches are, consistent with self-activation being dispensable for bistability (Sec.~\ref{sec:result1}).
Finally, $K_c$, $m$ and $\delta$ control the $S$-dependent repression of $H$, while $\gamma_S$ and $\beta_0$ shape the $S$ signal that drives the H trajectory and therefore affect the H peak-to-plateau ratio.

\begin{figure}[!htbp]
\centering
\includegraphics[width=0.90\textwidth]{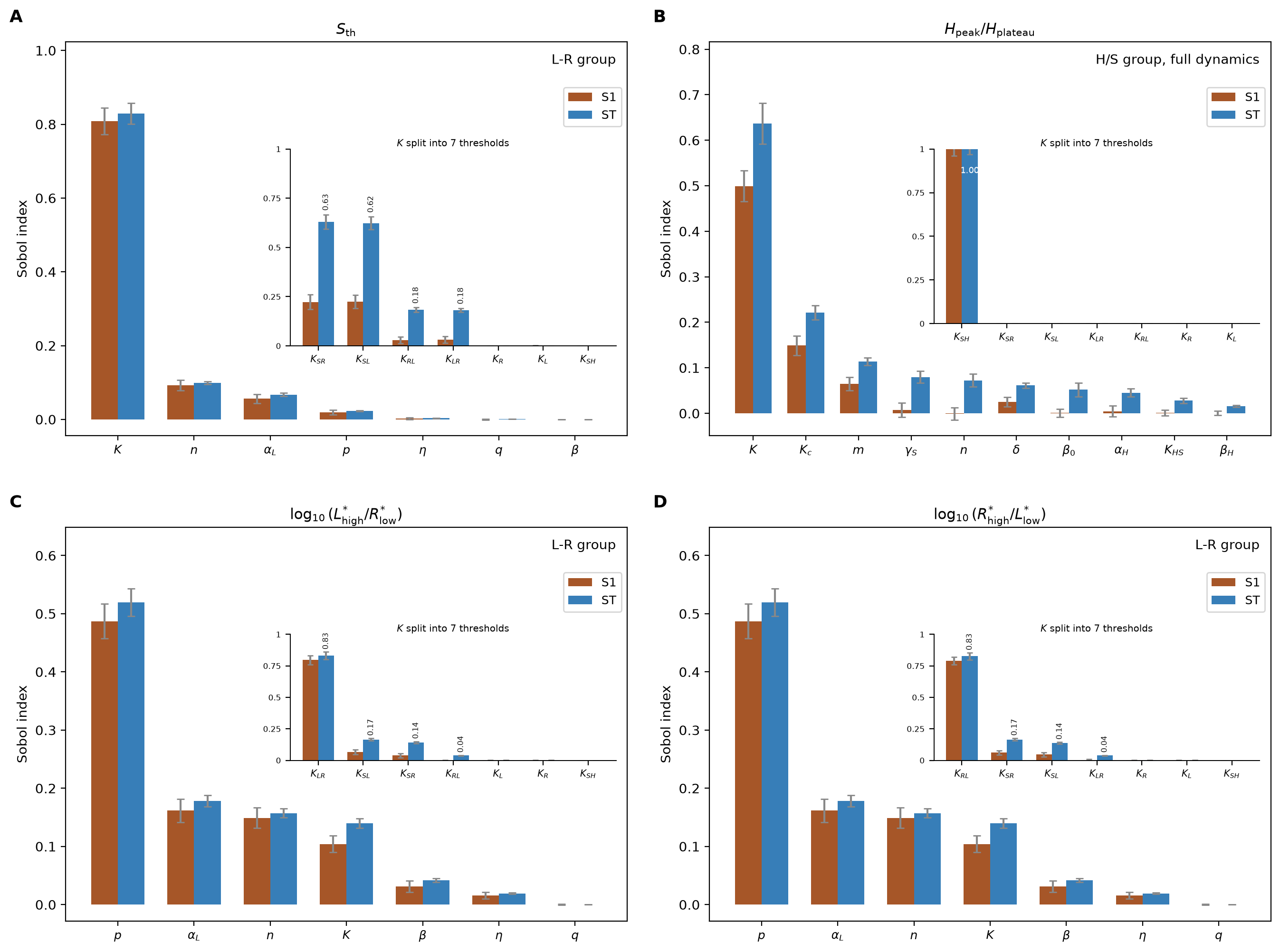}
\caption{Sobol global sensitivity of the four outputs.
	Brown bars give the first-order index $S1$ and blue bars the total-order index $ST$, both ordered by decreasing $ST$.
	Error bars show bootstrap $95\%$ confidence intervals.
	The Sobol indices were not truncated to $[0,1]$.
    In panel A, $S_{\mathrm{th}}$ is the bistability threshold of the L-R subsystem at fixed $S$; we evaluate its sensitivity by varying the seven retained L-R parameters $\alpha_L$, $\eta$, $\beta$, $K$, $p$, $n$ and $q$.
    In panel B, $H_{\mathrm{peak}}/H_{\mathrm{plateau}}$ is the peak-to-plateau ratio of the full $H(t)$ trajectory over the simulated period up to 36~h; we evaluate its sensitivity by varying the ten retained H/S parameters $\alpha_H$, $K$, $K_c$, $\delta$, $n$, $m$, $\beta_0$, $\beta_H$, $K_{HS}$ and $\gamma_S$.
	(C) $\log_{10}(L^{*}_{\mathrm{high}}/R^{*}_{\mathrm{low}})$ and
	(D) $\log_{10}(R^{*}_{\mathrm{high}}/L^{*}_{\mathrm{low}})$ are the two L-R expression ratios, each evaluated at a clamped size signal inside the bistable window, $S_{\mathrm{ref}}=\min(3S_{\mathrm{th}},0.6S_{\mathrm{hi}})$
	($\approx0.57$ at the reference values);
	the asterisk marks the corresponding steady state, and the two ratios are equal because the reference L-R	subsystem is symmetric about $L=R$.
	Ranges are $\pm50\%$ of the reference value for $\alpha_L$, $\eta$, $\beta$,
	$\alpha_H$ and $\delta$, 0.5--1.5 for the seven thresholds that share $K$,
	2--4 for $p$, $n$, $m$, 1--3 for $q$, and 1.0--2.0, 0.03--0.08, 0.5--1.3,0.3--0.8 and 0.05--0.20 for $K_c$, $\beta_0$, $\beta_H$, $K_{HS}$ and $\gamma_S$, respectively.
	Each design uses $N=4096$ Saltelli base samples without second-order terms,	that is $N(k+2)=36\,864$ model evaluations for $k=7$ and 49\,152 for $k=10$.
	The insets split the merged $K$ into $K_{SH}$, $K_{SL}$, $K_{SR}$, $K_{RL}$, $K_{LR}$, $K_L$ and $K_R$, perturbed independently with all other parameters at their reference values, on a 0--1 scale that is comparable across insets but not with the main panels: $K_{SR}$ and $K_{SL}$ dominate (A) ($ST=0.63$ and 0.62), $K_{SH}$ alone dominates (B) ($S1=ST=1.00$), $K_{LR}$ dominates (C) and $K_{RL}$ (D), and $K_L$ and $K_R$ stay below $10^{-3}$;splitting $K$ raises the summed total-order index in (A) to 1.62.}
\label{fig:sobol}
\end{figure}

\subsection{Data-constrained prediction of \textit{lag} expression}
\label{sec:result3}

\begin{figure}[!htbp]
\centering
\includegraphics[width=0.86\textwidth]{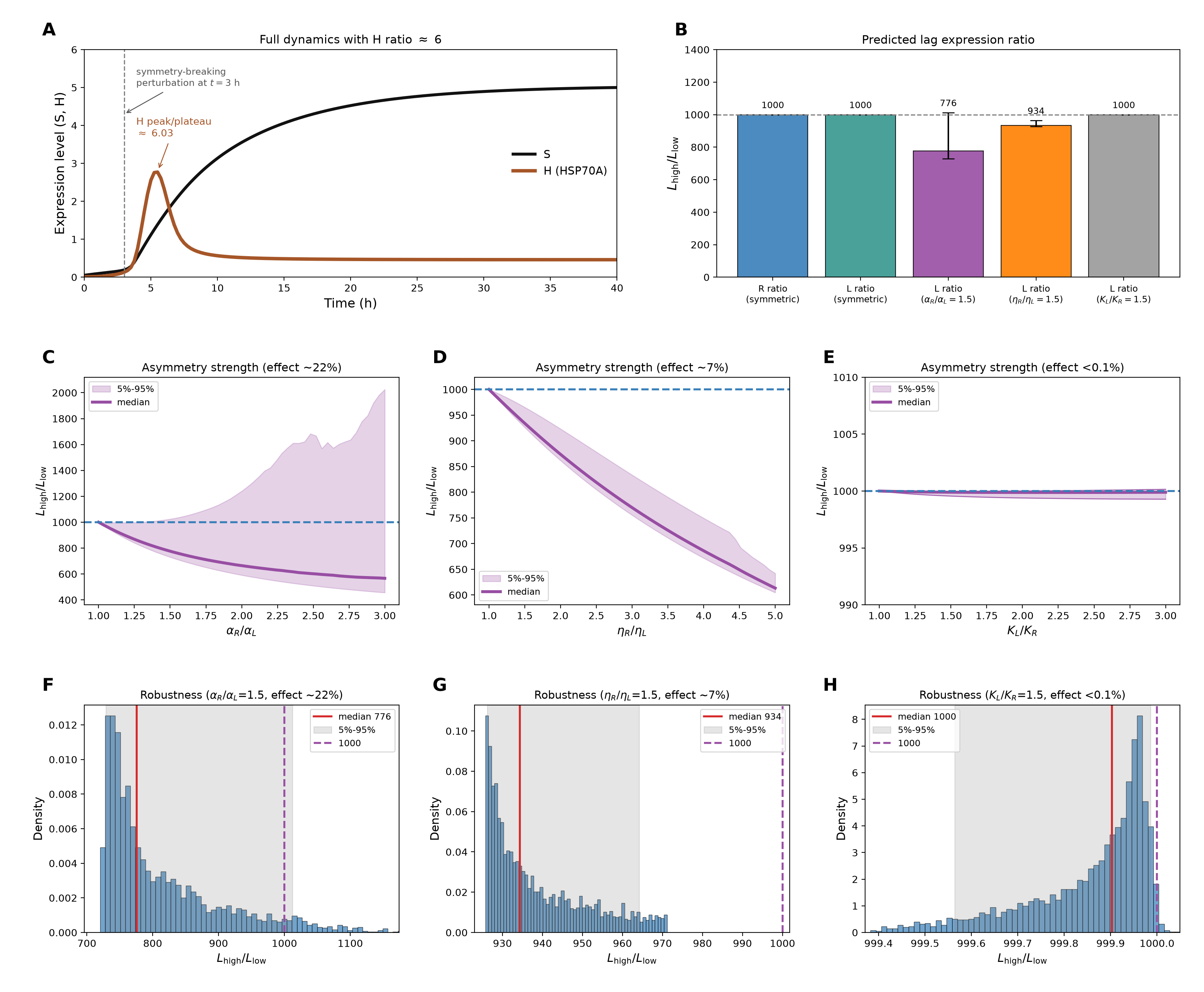}

\caption{Data-constrained prediction of the \textit{lag} expression ratio.
	The H/S parameters are held at the values calibrated in Sec.~\ref{sec:result3} ($K_c=2.275$, $\delta=7.60$, $m=3.87$, $\beta_0=0.0523$, $\beta_H=0.634$, $K_{HS}=0.264$, $\gamma_S=0.116$, $\alpha_H=3.5$, $n=3$) and the L/R parameters at the symmetric reference values of Table~\ref{tab:params} ($\alpha_L=\alpha_R=3.5$, $\eta_L=\eta_R=0.7$, $p=3$, $q=2$, $K=0.5$);
	$\beta_L=\beta_R=0.00394$ is re-tuned for every sample so that $R_{\mathrm{high}}/R_{\mathrm{low}}=1000$, and the steady states are evaluated at $S=5.04$.
	(A) Full dynamics with H ratio $\approx6$: $H$, $S$, $L$, $R$ from	$(0.01,0.05,0.05,0.05)$; a
	perturbation of $\pm0.02$ in $L$ and $R$ at $t=3$~h, when $S$ crosses the threshold, gives an $H$ peak-to-plateau ratio of 6.03.
	(B) Predicted lag expression ratio: medians with 5--95\% intervals from 3000 samples per mechanism at an asymmetry ratio of 1.5; both branches give 1000 in the symmetric case, whereas the $\alpha$, $\eta$ and $K$ asymmetries give 776 (729--1012), 934 (926--964) and 1000 (1000--1000); the two symmetric bars are deterministic single values and therefore carry no interval.
	(C-E) Asymmetry strength (effect $22\%$, $7\%$ and $<0.1\%$): predicted ratio against 51 evenly spaced strengths with 1000 samples each and common random numbers. 
	The vertical scales are independent: (C) and (D) span
	their own 5--95\% band, whereas (E) is fixed to 990--1010 ($\pm1\%$); the ratio falls to 567 and 613 at the upper ends of the $\alpha$ and $\eta$ scans.
	(F-H) Robustness at $\alpha_R/\alpha_L=1.5$, $\eta_R/\eta_L=1.5$ and $K_L/K_R=1.5$: distributions from the same 3000 samples, with medians in red, 5--95\% bands in grey and the symmetric value of 1000 shown dashed; the horizontal axes follow each distribution's 0.3--99.7 percentile range.}
\label{fig:result3}
\end{figure}

A direct experimental measurement of the \textit{lag} expression ratio is not yet available because \textit{lag} has not been fully cloned or sequenced \cite{kirk2001germ, kirk2005twelve}.
Two related quantities are known: the HSP70A-dependent module shows a transient rise and later decline during development, with a peak-to-plateau ratio of approximately 6 \cite{cheng2005hsp70a}, and \textit{regA} expression differs by approximately 1000-fold between the two cell types \cite{matt2018cell}.
Our aim in this section is to use these two observations as constraints and ask whether the model can predict the \textit{lag} expression ratio without fitting to \textit{lag} itself.
All degradation rates are fixed at $\gamma=0.35\,\mathrm{h^{-1}}$ from protein-turnover data.
Direct kinetic measurements are not available for most H/S parameters, so their values are calibrated rather than measured.
We adjust $K_c$, $\delta$, $m$, $\beta_0$, $\beta_H$, $K_{HS}$ and $\gamma_S$ so that the model reproduces an H peak-to-plateau ratio of approximately 6, $S$ crosses the L-R bistability threshold at approximately 3~h, consistent with the developmental timing of asymmetric cleavage, and all variables reach steady state.
Under synchronized light--dark conditions, one complete asexual cycle of \textit{Volvox carteri} is completed every 48~h \cite{kirk2001germ}.
We therefore integrate the full dynamics over a 40~h window, which lies within this cycle and is used to confirm that the trajectories have settled rather than as a measured developmental endpoint.
The H plateau is evaluated at 36~h within this window.
These constraints do not determine a unique parameter set; the values used here provide one calibrated solution that satisfies these conditions, with a calculated H ratio of 6.03 (Fig.~\ref{fig:result3}A).
The H/S-specific parameters adjusted here control the time courses of $S$ and $H$, including when $S$ reaches the L-R threshold, but they do not enter the L-R equations and therefore do not change $S_{\mathrm{th}}$ or the separation between the two L-R states.
The shared parameters $K$ and $n$ appear in both subsystems, and their contributions to $S_{\mathrm{th}}$ arise from the L-R terms, as confirmed by the dominant effects of $K_{SL}$ and $K_{SR}$ in the split-$K$ analysis.
These quantities are therefore controlled by the L-R parameters, consistent with the largely separate parameter sets identified in Sec.~\ref{sec:result2}.
We therefore hold the calibrated H/S subsystem fixed and vary the L-R parameters separately in the three asymmetry cases described below.
With the H/S subsystem fixed, we next use the L-R model to predict the \textit{lag} expression ratio under different parameter settings, taking the symmetric network as the baseline and focusing on the three asymmetric cases.

In the symmetric reference case ($\alpha_L=\alpha_R=3.5$, $\eta_L=\eta_R=0.7$ and equal half-saturation constants), $\beta_L=\beta_R$ is tuned so that $R_{\mathrm{high}}/R_{\mathrm{low}}=1000$.
Because the $L$ and $R$ equations are then identical, the same parameter set gives $L_{\mathrm{high}}/L_{\mathrm{low}}=1000$, so the model predicts a \textit{lag} ratio of 1000 in the symmetric case (Fig.~\ref{fig:result3}B).
To examine how a directional bias toward the somatic fate changes this prediction, we introduce three minimal asymmetries.
We considered three branch-specific mechanisms: a higher maximal $S$-activated production rate in the $R$ branch ($\alpha_R>\alpha_L$), stronger $R$ self-activation ($\eta_R>\eta_L$), and a lower half-saturation constant for $R$ self-activation ($K_R<K_L$).
These parameters can be assigned separately to $L$ and $R$, so each case creates a one-sided bias toward the somatic fate.
We did not vary $\beta_L$ and $\beta_R$ independently because they are used to impose the experimentally constrained $R_{\mathrm{high}}/R_{\mathrm{low}}=1000$.
We also did not use $n$ or $p$ because they are shared by both branches and control the steepness of the $S$-dependent and mutual-inhibition terms, rather than creating an $R$-specific bias.
Their selection is based on these branch-specific roles rather than on their exact Sobol ranking; in particular, $p$ has the largest Sobol index for the L-R expression ratio but is common to both branches and therefore cannot be used to construct a one-sided somatic bias.
The asymmetries are motivated by the observation that somatic cells are terminally differentiated and must sustain high \textit{regA} expression to repress reproductive functions, whereas germ cells retain reproductive potential; loss of \textit{regA} releases somatic cells from this committed state and allows them to resume division \cite{kirk2001germ, stark2001two, matt2018cell, hallmann2011evolution, kirk1999regA}.
Accordingly, the $R$-dominant somatic branch is assigned the more robustly maintained fate through a higher maximal $S$-activated production rate ($\alpha_R>\alpha_L$), stronger self-activation ($\eta_R>\eta_L$) or a lower self-activation half-saturation constant ($K_R<K_L$).
For each asymmetry the remaining L/R parameters are kept symmetric and $\beta_L=\beta_R$ is retuned so that $R_{\mathrm{high}}/R_{\mathrm{low}}=1000$ is maintained.

To compare the three mechanisms, we define asymmetry strength as the ratio of the corresponding branch-specific parameters: $\alpha_R/\alpha_L$, $\eta_R/\eta_L$ or $K_L/K_R$.
A value of $1$ corresponds to the symmetric case, and values above $1$ indicate a stronger bias toward the somatic branch.
At an asymmetry strength of $1.5$, the predicted median \textit{lag} ratio is 776 for the $\alpha$ asymmetry (5--95\% interval 729--1012), 934 for the $\eta$ asymmetry (926--964), and 1000 for the $K$ asymmetry (1000--1000) (Fig.~\ref{fig:result3}B).
Relative to the symmetric prediction of 1000, these correspond to reductions of approximately $22\%$ for $\alpha$, $7\%$ for $\eta$ and less than $0.1\%$ for $K$.
Over the scanned ranges, both the $\alpha$ and $\eta$ asymmetries reduce the predicted \textit{lag} ratio monotonically, whereas the $K$ asymmetry has a negligible effect (Fig.~\ref{fig:result3}C--E).
The $\alpha$ mechanism produces a steeper decrease and a much wider 5--95\% interval, whereas the $\eta$ mechanism has a smaller effect and a narrower interval.

The three parameters give different predictions because $\alpha$ and $\eta$ enter the production terms, whereas $K$ sets the concentration at which self-activation saturates; the $\alpha$ and $\eta$ asymmetries are also imposed under different constraints.
Once $S$ is large enough, self-activation is close to saturation in the high-expression state and negligible in the low-expression state, while mutual inhibition is weak in the high state.
The synthesis rate in the high state is then approximately $\beta+\alpha_i+\eta_i$, whereas the low-state synthesis rate is dominated by $\beta$.
Because both genes have the same degradation rate, the high/low ratio of gene $i$ is approximately $(\beta+\alpha_i+\eta_i)/\beta$, with $\gamma$ cancelling.
Imposing $R_{\mathrm{high}}/R_{\mathrm{low}}=1000$ through $\beta_L=\beta_R$ gives $\beta\approx(\alpha_R+\eta_R)/999$, and hence
\begin{equation}
	\frac{L_{\mathrm{high}}}{L_{\mathrm{low}}} \;\approx\;
	1 + 999\,\frac{\alpha_L+\eta_L}{\alpha_R+\eta_R}.
	\label{eq:L_ratio_asym}
\end{equation}

This expression treats $\alpha$ and $\eta$ in the same way, so their different effects come from how the asymmetries are imposed.
In the $\alpha$ case, $\alpha_L$ and $\alpha_R$ are shifted in opposite directions while $\alpha_L+\alpha_R$ is fixed, so both the numerator and denominator of $(\alpha_L+\eta_L)/(\alpha_R+\eta_R)$ change.
At the reference values, $\alpha_R/\alpha_L=1.5$ gives $(\alpha_L+\eta_L)/(\alpha_R+\eta_R)\approx3.5/4.9\approx0.71$, which predicts an $L$ ratio near 710, close to the simulated median of 776 (Fig.~\ref{fig:result3}B).
The remaining difference comes mainly from residual cross-inhibition and self-activation in the low-expression state, which are neglected in the leading-order expression.
In the $\eta$ case, only $\eta_R$ increases, while $\alpha$ remains unchanged.
Since $\alpha=3.5$ is much larger than $\eta=0.7$ at the reference point, increasing $\eta_R$ to $1.05$ changes the total production only from $4.2$ to $4.55$.
This gives a ratio of totals near $4.2/4.55\approx0.92$ and an $L$ ratio near 920, close to the simulated median of 934.
The smaller effect of the $\eta$ asymmetry therefore reflects the small fractional change in total production rather than a smaller weight of $\eta$ in Eq.~(\ref{eq:L_ratio_asym}).
The two settings also represent different biological possibilities: shifting $\alpha_L$ and $\alpha_R$ in opposite directions corresponds to a conserved total $S$-activated production budget, whereas increasing $\eta_R$ alone corresponds to additional self-reinforcement in the somatic branch.
Changing the normalization of the $\eta$ asymmetry does not alter the conclusion; keeping $\eta_L+\eta_R$ fixed instead changes the simulated median only from 934 to 947, still well above the $\alpha$ case.
The $K$ asymmetry has almost no effect on the predicted ratio because $K$ controls the concentration at which self-activation saturates, whereas the high/low ratio is set by the saturated production rates $\alpha$ and $\eta$.
A $K$ asymmetry therefore shifts the bistable branch but leaves the predicted \textit{lag} ratio close to 1000.

Finally, we assessed the robustness of the predicted ordering to parameter uncertainty (Fig.~\ref{fig:result3}F--H).
The predicted $K$ ratio remains concentrated near the symmetric value of 1000, the $\eta$ distribution lies entirely above 925, and 88\% of the $\alpha$ samples fall below the $\eta$ median.
These results reflect the different roles of the three parameters: $K$ mainly shifts the bistable branch without changing the saturated high/low ratio, whereas $\alpha$ and $\eta$ change the total production, with $\eta$ producing the smaller change under the imposed constraint.
The predicted ratios therefore remain ordered as $\alpha<\eta<K$ across the sampled parameter sets, indicating that this ranking is not driven by a particular parameter choice.
In (B) the first two bars are deterministic single values and therefore carry no interval. In (C)--(E) the grids have spacings of 0.04 in (C) and (E) and 0.08 in (D), and therefore do not contain 1.5 exactly, so the value at 1.5 is taken from the separate calculation shown in (B); (C) and (D) span the 5--95\% band of their own data, whereas (E) is fixed to 990--1010 ($\pm1\%$) so that the nearly flat $K$ dependence is not exaggerated by autoscaling. In (F)--(H) the horizontal axes follow each distribution's 0.3--99.7 percentile range, that of (G) being extended so that the symmetric value of 1000 remains visible; only an asymmetry in the $S$-activated production rate materially lowers the predicted \textit{lag} ratio, whereas asymmetries in self-activation strength or in the self-activation threshold of this size are buffered by the mutual inhibition.

\section{Discussion}

We developed a four-variable mathematical model of \textit{Volvox carteri} germ-soma differentiation to examine how cell size controls fate specification and to predict the \textit{lag} expression ratio.
Bifurcation analysis, global sensitivity analysis and experimentally constrained simulations identified three main results.
First, the cell-size signal $S$ is required for bistability: the network cannot produce two stable cell fates without a sufficiently strong size signal.
Second, mutual inhibition is required for bistability, whereas self-activation is not required and does not change the number of stable differentiated states.
Self-activation does, however, accelerate commitment and, when asymmetric, changes the predicted \textit{lag} expression ratio.
Third, different branch-specific asymmetries produce distinct predictions for the \textit{lag} ratio even when all cases reproduce the same \textit{regA} ratio.
The main finding of this study is that $S$ is a necessary condition for germ-soma commitment, rather than merely a modulator of an otherwise bistable switch.
Mutual inhibition can support bistability in an isolated toggle switch, but in the full model the network does not become bistable without $S$.
This gives the experimentally observable size cue a central role that earlier models of toggle-like fate switches have largely left implicit.

The main finding of this study is that the cell-size signal $S$ is a necessary condition for germ-soma commitment, rather than merely a modulator of an otherwise bistable switch.
Mutual inhibition can support bistability in an isolated toggle switch, but in the full model the network does not become bistable without $S$.
This gives the experimentally observable size cue a central role that earlier models of toggle-like fate switches have largely left implicit.
A second finding is that the predicted \textit{lag} expression ratio depends on which L-R parameter is asymmetric, even when all cases reproduce the same \textit{regA} ratio of approximately 1000.
At an asymmetry ratio of 1.5, an $\alpha$ asymmetry, in which the somatic branch has a stronger maximal $S$-activated production rate, lowers the predicted \textit{lag} ratio to a median of 776.
An $\eta$ asymmetry gives a median of approximately 934, whereas a $K$ asymmetry leaves the prediction close to 1000.
This provides a quantitative and falsifiable prediction that can be tested once \textit{lag} has been cloned and its expression measured.
It also connects the established terminal commitment of somatic cells \cite{kirk2001germ, stark2001two, matt2018cell} to a specific molecular mechanism.

The L-R core of the model is the classical genetic toggle switch \cite{gardner2000construction}, but here it is placed downstream of the cell-size signal $S$.
For each fate determinant, the production term is written as the product of an $S$-dependent factor, a cross-repression factor and the term $\alpha_i+\eta_iF_i(X_i)$, where $F_i(X_i)$ describes self-activation.
This multiplicative form means that self-activation can strengthen $X_i$ production only when $S$ is present and repression by the opposite gene is weak.
It therefore allows self-activation to reinforce an already selected fate without allowing it to initiate that fate in the absence of $S$.
If self-activation were added directly to the production term, $L$ or $R$ could maintain itself even when $S=0$ or when the opposite determinant was strongly expressed.
Such an additive form would weaken the dependence on the size signal and blur the mutually exclusive nature of the two fates.
We include self-activation as a possible positive-feedback mechanism, although it is not required for bistability.
Removing self-activation leaves two stable states but shifts the threshold from $S_{\mathrm{th}}\approx0.19$ to approximately 0.20 and slows the approach to the committed states.
When self-activation is asymmetric, it also changes the predicted \textit{lag} ratio.
This model structure is consistent with lineage-commitment models in bipotent progenitors \cite{huang2007bifurcation} and with the view that differentiation hierarchies and irreversibility depend on the structure of the regulatory network \cite{nakamura2024evolution}.

The model has several limitations.
Most kinetic parameters for \textit{Volvox} have not been measured directly, so their values are based on data from closely related algae or on ranges used in earlier gene-regulatory models.
Only the degradation rate $\gamma$ is informed by protein-turnover measurements \cite{mastrobuoni2012proteome, martin2012proteome, raynaud2007mca1}.
We use $\gamma=0.35\,\mathrm{h^{-1}}$ for short-lived regulatory proteins, which is much faster than the bulk-proteome estimate of approximately $0.015\,\mathrm{h^{-1}}$.
The variables $S$ and $H$ represent effective modules rather than measured molecular concentrations, so the estimated threshold $S_{\mathrm{th}}\approx0.19$ cannot yet be linked to a specific cell size.
The predicted \textit{lag} ratio remains untested because \textit{lag} has not been cloned or sequenced \cite{kirk2001germ, kirk2005twelve}; the prediction can be examined once suitable expression measurements become available.
The model is also deterministic and describes a single cell, so it does not include stochastic expression noise \cite{lei2009stochastic, jia2017simplification} or the sequence of asymmetric divisions that builds the colony.
These processes are important directions for extending the model.

Future work should address these gaps.
The most direct test is to clone and sequence \textit{lag} and measure its expression in germ and somatic cells.
The model predicts a \textit{lag} ratio of approximately 776 for an $\alpha$ asymmetry, 934 for an $\eta$ asymmetry and near 1000 for a $K$ asymmetry at an asymmetry ratio of 1.5.
A measured ratio close to one of these values would help identify which mechanism is more consistent with the observed somatic-cell commitment.
Direct measurements of HSP70A dynamics and of \textit{lag} and \textit{regA} turnover would also provide stronger constraints on the kinetic parameters.
In addition, $S$ and $H$ should be linked to measurable cell-size and molecular readouts so that the predicted threshold can be expressed in physical units.
Finally, stochastic and multicellular models that include successive asymmetric divisions would allow the effects of expression noise and lineage history to be examined.
These extensions would turn the present theoretical predictions into more directly testable hypotheses.

\section*{Acknowledgments}

This work was supported by the National Natural Science Foundation of China (grant numbers 12401644 and 12571538).

\end{document}